\documentclass[prb,twocolumn,amsmath,amssymb,superscriptaddress,floatfix,aps]{revtex4-2}
\usepackage{graphicx,subfigure,amsmath,bm,colordvi,times}
\usepackage{mathtools}
\usepackage[usenames]{color}
\usepackage{multirow}
\usepackage{epstopdf}
\usepackage[utf8]{inputenc}
\usepackage[english]{babel}
\usepackage{amsfonts}
\usepackage{amssymb}
\usepackage{float}
\usepackage{hyperref}
\usepackage{braket}
\usepackage[normalem]{ulem}
\usepackage{comment}

\usepackage[normalem]{ulem}
\hypersetup{pdfborder=0 0 0,colorlinks=true,citecolor=blue,linkcolor=blue}

\begin{document}

\title{Vortices in dipolar condensates of interlayer excitons}
\author{Sara Conti}
\email{sara.conti@uantwerpen.be}
\affiliation{
Department of Physics, University of Antwerp, Groenenborgerlaan 171, 2020 Antwerp, Belgium}

\author{Andrey Chaves}
\affiliation{Departamento de F\'isica, Universidade Federal do Cear\'a, 60455-900 Fortaleza, Cear\'a, Brazil}

\author{Luis A. Pe\~{n}a Ardila}
\affiliation{
Dipartimento di Fisica, Universit\`{a} di Trieste, Strada Costiera 11, I-34151 Trieste, Italy}

\author{David Neilson}
\affiliation{
Department of Physics, University of Antwerp, Groenenborgerlaan 171, 2020 Antwerp, Belgium}

\author{Milorad V. Milo\v{s}evi\'c}
\email{milorad.milosevic@uantwerpen.be}
\affiliation{
Department of Physics, University of Antwerp, Groenenborgerlaan 171, 2020 Antwerp, Belgium}

\begin{abstract}
Recently observed signatures of Bose-Einstein condensation and superfluidity of dipolar excitons have drawn enormous attention to excitonic semiconductor bilayers. In superfluids, stabilization and observation of vortex matter is usually a decisive proof of coherent condensation order. However to date, the vortex behavior in a 2D excitonic system with aligned dipole-like interactions that are long-range and everywhere repulsive has not been addressed. We here provide a theoretical description of the vortex characteristics, interaction, and lattices in a dipolar exciton superfluid, solving the corresponding Gross-Pitaevskii equation, while varying the exciton dipole moments and the exciton density -- both tunable in the experiment, by interlayer separation and gating, respectively. We draw particular attention to the appearance of a maximum in the density redistribution around the edge of each vortex, in the phase-space region where the dipole interactions are particularly strong, and where a transition to an incompressible exciton supersolid is expected. 
\end{abstract}
\maketitle

\section{Introduction}

A number of observations of signatures of Bose-Einstein condensation \cite{Burg2018,Ma2021} and superfluidity \cite{Nguyen2023, Cutshall2025} of dipolar excitons have drawn enormous attention to excitonic bilayer solid-state semiconductor systems in which the electrons and holes are kept separate in adjacent 2D conducting layers \cite{Neilson2024}.
The exciton bilayer is an appealing physical system because of the relatively high transition temperatures of these condensed phases and the promise of integrable low-energy electronic devices~\cite{Fogler2014,Conti2021b}. 
In addition, a transition to an incompressible exciton supersolid has recently been predicted~\cite{Conti2023}, depending on the exciton density and exciton dipole moment in the hosting solid-state system. 

Unambiguously identifying the superfluid and supersolid phases remains challenging, mainly because excitons are neutral quasiparticles. 
In the absence of a Meissner effect for a neutral system, the existence of vortices is among the definitive proofs to identify superfluidity~\cite{Onsager1949,Rayfield1963}.  
Persistent currents in superfluids are intimately related to the existence of quantized vortices which are localized phase singularities with an integer topological charge \cite{Gross1961, Pitaevskii1961}. 
Superfluid vortices have been extensively studied and observed in ultracold gases with short-ranged contact interactions between the atoms \cite{AboShaeer2001}, and more recently in ultracold atoms with large magnetic dipolar moments~\cite{Neely2010}. 
In dipolar ultracold gases, the properties of the vortices depend on the balance between the repulsive and attractive atomic interactions \cite{Mulkerin2013}. 
The vortices are also predicted to exhibit peculiar changes in shape and critical angular velocity if the system enters a supersolid phase \cite{Gallemi2020}.
The first observation of vortices in dipolar condensates \cite{Klaus2022}, and more recently in dipolar supersolids \cite{Casotti2024}, have also been reported.

We here characterize the existence and properties of vortices in the exciton bilayer system with the aim of using vortices as an experimental probe to establish the existence of the quantum condensed phases. 
Vortices could be generated experimentally in this system by optically imprinting the vortex phase pattern~\cite{Delpace2022} or by direct transfer of orbital angular momentum from a superchiral laser beam source~\cite{Yokoyama2020}.

Unlike dipolar ultracold gas condensates, which can in general involve both repulsive and attractive interactions, in the exciton system the dipolar interactions are always repulsive since the exciton dipoles are aligned perpendicular to the layers. 
The purely repulsive interactions suggest that the formation and properties of the vortices for the exciton system can differ fundamentally from ultracold gases, leading for example to unique behavior of the vortices near the supersolid transition. 
In ultracold gases, the short-range attractive physics is crucial to achieve a supersolid, while in the exciton system the incompressible supersolid can be driven by purely repulsive interactions \cite{Conti2023}.
Understanding this new behavior is among the objectives of the present work.

The paper is organized as follows.
In Sec.\ \ref{Sec:Theoretical} we extend the Gross-Pitaevskii formalism for dipolar gases to account for the exciton-exciton interaction in an electron-hole bilayer system. 
We introduce the model to determine the single-vortex properties (results presented in Sec.\ \ref{Sec:SingleVortex}), the vortex-vortex interaction (Sec.\ \ref{Sec:VortexInt}) and the vortex lattice (Sec.\ \ref{Sec:Lattice}) in the superfluid state of the dipolar excitonic system.
Our main points and conclusions are then summarized in Sec.\ \ref{Sec:Conclusions}.

\section{Theoretical formalism}
\label{Sec:Theoretical}

At very low densities the interlayer exciton gas behaves as a dilute Bose gas and the Bose-Einstein condensate (BEC) \cite{Perali2013} can be described by a macroscopic wave function (an ‘order parameter’) $\Psi(\mathbf{r})$. 

We capture the dynamics of the exciton BEC by the time-independent Gross–Pitaevskii equation \cite{Gross1961, Pitaevskii1961} for the condensate wave function $\Psi(\mathbf{r},t)$, extended to include the dipolar interaction between aligned excitons. We modify the standard Gross–Pitaevskii equation for dipolar gases \cite{Goral2000} by eliminating both the external harmonic confinement and the attractive contact interaction. This leaves the exciton-exciton interaction as the sole interaction. 

We consider exciton dipoles aligned perpendicularly to the bilayer, such that the exciton-exciton interaction $V_{\text{XX}}$ is given by
\begin{equation}
    V_{\text{XX}}(r)=\frac{2 e^2}{4 \pi \epsilon} \left(\frac{1}{r}-\frac{1}{\sqrt{r^2+d^2}} \right)\ ,
    \label{Eq:DipolarPotential}
\end{equation} 
where $\mathbf{r}$ is the in-plane inter-particle vector. $V_{\text{XX}}(r)$ contains the four Coulomb interactions acting between the constituent electrons and holes separated by an interlayer distance $d$ \cite{Conti2023}. For $d$ smaller than the inter-particle distance, $V_{\text{XX}}$ is a pure dipolar interaction. 
The electric dipole moment is tunable since it is proportional to the electron-hole layer separation.

The stationary solutions take the form $\Psi(\mathbf{r},t)\!\! = \!\!\Psi(\mathbf{r})\, \mathrm{e}^{-i\mu t/\hbar}$, with $\mu$ being the chemical potential, and satisfies the stationary dipolar Gross–Pitaevskii equation, 
\begin{equation}
-\frac{\hbar^2}{2M} \nabla^2 \; \Psi(\mathbf{r}) +  \phi_{\text{XX}}(\mathbf{r})\Psi(\mathbf{r})=\mu \Psi(\mathbf{r})\ ,
\label{Eq:DGPEquation}
\end{equation}
where $M=m_e^*+m_h^*$ is the exciton mass with effective electron and hole masses taken to be equal $m_e^*=m_h^*$.
The exciton mean-field potential $\phi_{\text{XX}}$ is,
\begin{equation}
    \phi_{\text{XX}}(\mathbf{r})=\int V_{\text{XX}}(|\mathbf{r}-\mathbf{r'}|) \left| \Psi(\mathbf{r'})\right|^2 d\mathbf{r'}\ ,
    \label{Eq:DipolarEnergy}
\end{equation}
where $n(\mathbf{r})=\left| \Psi(\mathbf{r'})\right|^2 $ is the 2D particle density distribution.

We solve Eq.\ \eqref{Eq:DGPEquation} by evolving it in imaginary time \cite{Haas2018}, starting from a spatially randomized order parameter.
The potential $\phi_{\text{XX}}(\mathbf{r})$ for the dipole interaction is non-local, thus making the analytical treatment significantly different from the case of contact interactions \cite{Haas2018}. 
We evaluate Eq.\ \eqref{Eq:DipolarEnergy} in Fourier space and use the convolution theorem
$\phi_{\text{XX}}(\mathbf{r})=\mathcal{F}^{-1}[V_{\text{XX}}(k)n(\mathbf{k})] $~\cite{Fischer2006}.
All simulations are performed on a $256^2$ grid covering the sample spatial area, with Dirichlet boundary conditions. 
Solutions are accepted as converged when the relative change in energy in consecutive time steps decreases below $10^{-11}$.

\begin{figure}[t]
\centering
\includegraphics[width=\columnwidth]{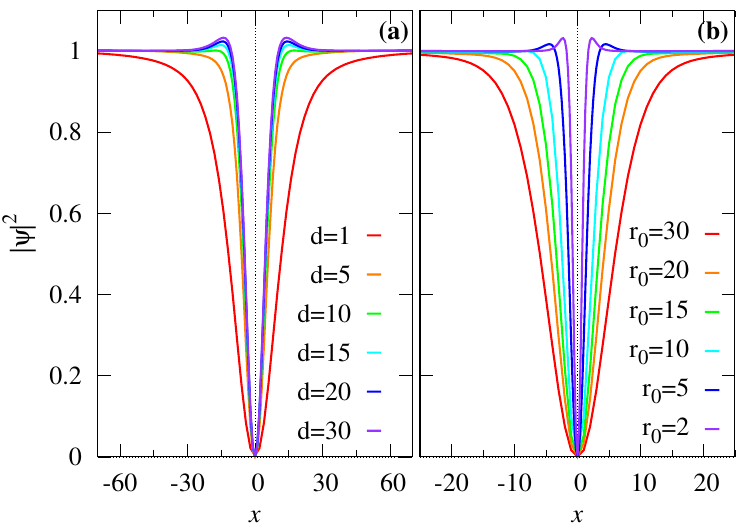}
\caption{(a) Cross section of vortex density profile $|\psi(r)|^2$ along the $x$-axis at a fixed density, given by the average inter-particle distance $r_0=30$, for different interlayer distances.
(b) The cross section of vortex density profile $|\psi(r)|^2$ at a fixed interlayer distance $d=5$, for different exciton densities. 
}
\label{fig:vortex_d}
\end{figure}

In what follows, we use the effective Bohr radius $a_B=\hbar^2 4 \pi \epsilon/e^2 m_e^*$ as a unit length scale, with dielectric constant $\epsilon$ for a bilayer semiconductor system. The energy is expressed in effective Rydberg (Ry$^*$).

\section{Results}
\subsection{Single-vortex properties}
\label{Sec:SingleVortex}

We start by considering a single-vortex solution in the 2D homogeneous system, i.e. a single vortex pinned at the origin, with an associated order parameter:
\begin{equation}
\Psi(\mathbf{r}) = \sqrt{N}\psi(r)e^{i\ell\theta(\mathbf{r})}\,.
\label{Eq:1Vwavefunction}
\end{equation}
where $N$ is the number of particles and $\ell$ the vorticity, i.e., the quanta of phase circulation carried by the vortex. 
The phase $\theta$ is set as $\theta(\mathbf{r})=\arctan(y/x)$ with $x$ and $y$ the Cartesian coordinate of $\mathbf{r}$.
Equation~\eqref{Eq:DGPEquation} then becomes
\begin{align}
-\frac{\hbar^2}{2M} \nabla^2 \psi(r) +\frac{\ell^2 \hbar^2}{2M} \frac{1}{r^2}\psi(r)
+\phi_{\text{XX}}(r) \psi(r) =\mu  \psi(r)\ .
\label{Eq:DGPEquation1}
\end{align}

\begin{figure}[t]
\centering
\includegraphics[width=\columnwidth]{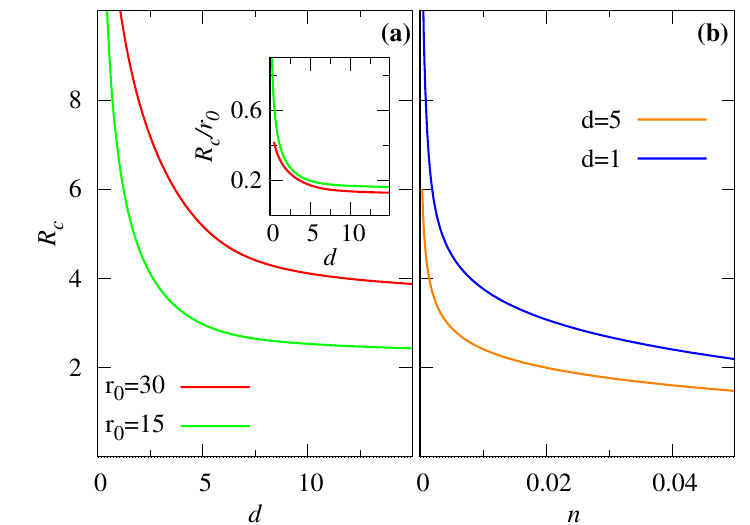}
\caption{(a) Vortex core radius $R_c$ as a function of interlayer spacing $d$, for $r_0=15$ and $30$. Inset shows $R_c/r_0$. (b) $R_c$ as a function of density, for $d=1$ and $5$.}
\label{fig:vortexsize}
\end{figure}

\begin{figure*}[ht]
\centering
\includegraphics[ width=0.7\textwidth]{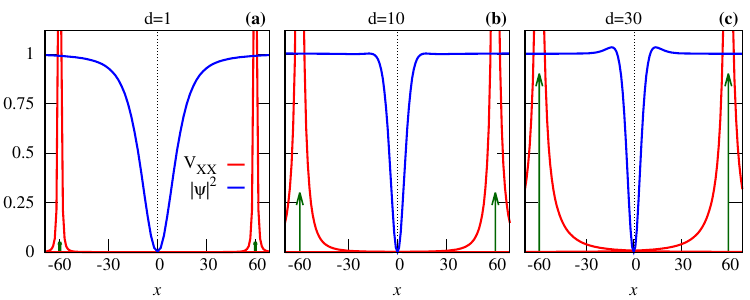}
\caption{Cross section of vortex density profile $|\psi(r)|^2$ (blue) normalized to the homogeneous density, and the exciton-exciton interactions $V_{\text{XX}}(r)$ in Ry$^*$ (red).
The green arrows are proportional to the dipole moments of the neighboring excitons. The density $n$ corresponds to $r_0=30$. Panels (a)-(c) are for interlayer distances $d=1$, $10$, and $30$, respectively.}
\label{fig:wings}
\end{figure*}
\begin{figure}[b]
    \centering
\includegraphics[width=\columnwidth]{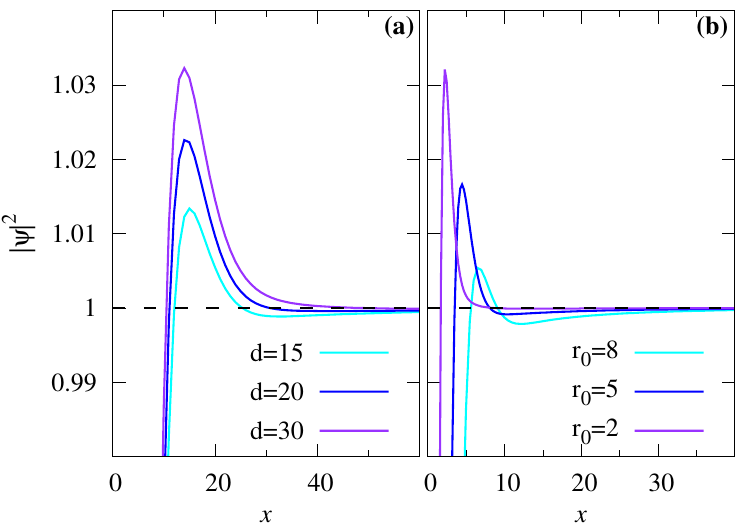}
    \caption{Detail of the vortex profile density pileup at (a) a fixed density, 
 $r_0=30$, and (b) a fixed interlayer distance, $d=5$.}
    \label{fig:peakzoom}
\end{figure}
\begin{figure}[b]
\centering
\includegraphics[trim={0.5cm 0.4cm 0.0cm 0.3cm},clip=true,width=\columnwidth]{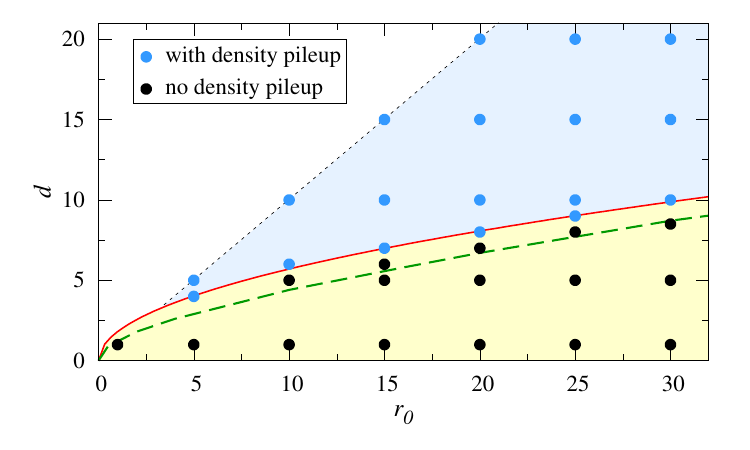}
\caption{Zero temperature phase diagram as function of the layer separation $d$ and average inter-particle distance $r_0$. 
The dotted black line is $r_0=d$. 
Blue (black) dots stand for superfluid systems with vortex profiles with (without) a density pileup peak. 
The red line is the dipolar coupling parametric threshold with $D=13$.
The green dashed line is the superfluid to supersolid transition in Ref.\ \onlinecite{Conti2023}.
}
\label{fig:phasediagram}
\end{figure}
\begin{figure}[t]
    \centering
    \includegraphics[trim={0.4cm 0.5cm 0.4cm 0.0cm},clip=true,width=0.95\columnwidth]{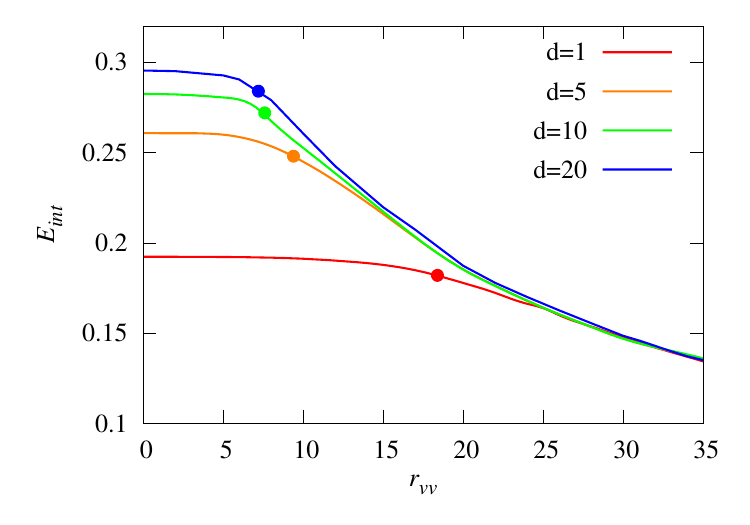}
    \caption{Interaction energy $E_{int}$ of two vortices in the exciton superfluid as a function of vortex separation $r_{vv}$, for different interlayer distances $d$ and a fixed density $r_0=30$. The dots mark where $r_{vv}=2 R_c$ in the different cases considered. }
    \label{fig:energy}
\end{figure}

Figure \ref{fig:vortex_d} presents the calculated spatial dependence of the order parameter around a vortex with vorticity $\ell=1$ on the experimental system parameters, namely the layer separation $d$ and the density $n \equiv 1/(\pi r_0^2)$.  
Figure \ref{fig:vortex_d}(a) shows, for a fixed density, the vortex profile for different interlayer distances $d$, while Fig.\ \ref{fig:vortex_d}(b) shows the vortex profile for fixed $d$, but with density varied.

The radius of the vortex core $R_c$, taken as the half-width at half maximum of the superfluid density distribution at the vortex, tends to decrease with increasing interlayer distance $d$ (Fig.\ \ref{fig:vortexsize}(a)) or increasing density $n$ (Fig.\ \ref{fig:vortexsize}(b)).
We recall that the healing length characterizing the vortex core radius is defined as $\xi=\sqrt{\frac{\hbar^2}{2M \mu}}$ \cite{Martin2017}.
As $\phi_{\text{XX}}$ becomes stronger by increasing the dipolar moment proportional to $d^2$ or by reducing $r_0$, $\xi$ tends to decrease.
Eventually, however, $\xi$ and hence $R_c$ saturate at a small value much less than $r_0$ [see inset Fig.\ \ref{fig:vortexsize}(a)]. 
A similar saturation has been reported in cold atoms at the supersolid transition, which is the point where the vortex core is no longer characterized by inter-particle interactions but instead is determined by the crystal structure \cite{Gallemi2020}.

Figure \ref{fig:vortex_d} shows that above a certain value of $d$ or $n$, a peak appears in the vortex profile near the edge of the vortex, signaling an accumulation of superfluid density at the edge of the vortex.
The pileup peak is a feature of repulsive boson systems and has also been reported in dipolar cold atoms when the repulsion is strong \cite{Ancilotto2021}.
Increasing the layer separation $d$ (or decreasing $r_0$) has the effect of extending the effective range of the neighboring interactions $V_{\text{XX}}$ (solid red lines in Figure \ref{fig:wings}).
At the same time, the vortex radius $R_c$ shrinks, but by Fig.\ \ref{fig:wings}(b) it has saturated (solid blue lines). 
Comparing panels (a) and (b) of Fig.\ \ref{fig:wings} shows that the pileup at the vortex edge starts when the vortex radius saturates and the potentials $V_{\text{XX}}$ from the neighboring excitons are sufficiently strong that they are still significant at the edge of the vortex profile.
It is the competition between the centrifugal force, which is pushing the exciton out of the vortex core, and the $V_{\text{XX}}$, which is pushing back the excitons, that causes the density pileup at the edges.

Figure \ref{fig:peakzoom} shows that the maximum of the pileup peak depends on the strength of the exciton-exciton repulsion, and the position of the peak is fixed to the edge of the vortex. 
The dependence of the peak position on $r_0$ [Fig.\ \ref{fig:peakzoom}(b)] is similar to the one reported in dipolar cold atoms \cite{Ancilotto2021}.  
However, in our exciton system, we do not see the subsequent density oscillations accompanying the peak observed in cold atoms.  
These oscillations have been associated with a low-lying roton excitation that mixes with the ground state. 
The absence of oscillations in our system suggests that there is no roton excitation in this system.  
This is consistent with Martin {\it et al.}~\cite{Martin2017} who found for 2D dipolar gases that the roton instability is induced by either an attractive contact interaction or by
the attractive part of the dipolar interaction, and this is only present when the dipoles are not aligned perpendicular to the layer.

Figure \ref{fig:phasediagram} maps out the phase diagram of our system. 
The black dots mark the systems where the vortex profile has no pileup peak, and the blue dots the systems where there is a pileup peak.
The onset of the pileup peak closely follows the dipolar coupling parametric threshold (red line) $D=13$, with $D=\frac{e^2 d^2}{4 \pi \epsilon} \frac{2M}{\hbar^2} \frac{1}{r_0}=\frac{4d^2}{a_Br_0}$, namely the ratio of the dipolar energy to the kinetic energy \cite{Boning2011}.
The green dashed line marks the superfluid-to-incompressible supersolid transition driven by the strong exciton-exciton repulsion \cite{Conti2023}.  
Remarkably, the first appearance of the blue dots lies very close to this predicted transition, with $D=10$ strongly suggesting that the appearance of the pileup peak is a signature of the solidification.
This is consistent with our picture that the driving process for the density pileup at the edge of the vortices is closely related to the driving process for the localization of the excitons in the supersolid. 
Both occur when the exciton-exciton repulsion is sufficiently strong that its effective range becomes comparable to the inter-particle distance.

\subsection{Vortex-vortex interaction}
\label{Sec:VortexInt}

We determine the vortex-vortex interaction by pinning two vortices at a desired distance from each other and evaluating their interaction energy \cite{Mulkerin2013}. 
The order parameter in that case is given by:
\begin{equation}
\Psi(\mathbf{r}) = \sqrt{N}\psi(r)e^{i\ell_1\theta_1(\mathbf{r})}e^{i\ell_2\theta_2(\mathbf{r})}.
\label{Eq:2Vwavefunction}
\end{equation}
The phase distribution around the vortices is set as $\theta_i(\mathbf{r})=\arctan(y-y_i/x-x_i)$, with $r_i=(x_i , y_i)$ the location of the $i$-th vortex.
Equation \eqref{Eq:DGPEquation}, in case of equal vorticities $\ell_1=\ell_2=\ell$, then becomes \cite{Dantas2015}:
\begin{align}
-\frac{\hbar^2}{2M}\! \left[\nabla^2 \!-\ell^2 (X^2\!+\! 
Y^2)\right]\! \psi(r) 
+\phi_{\text{XX}}(r) \psi(r) =\mu  \psi(r),
\label{Eq:DGPEquationR}
\end{align}
with
\begin{equation}
    X=\frac{x-x_1}{r_1^2}+\frac{x-x_2}{r_2^2}\;,
    \qquad 
    Y=\frac{y-y_1}{r_1^2}+\frac{y-y_2}{r_2^2}\; .
\end{equation}

The energy change $E$ for a condensate with vortices is:
\begin{equation}
    E_j=\int d\mathbf{r} \left[
    \frac{\hbar^2}{2M}|\nabla \Psi_j(\mathbf{r})|^2 +\frac{1}{2}\phi_{\text{XX}}(\mathbf{r})|\Psi_j(\mathbf{r})|^2 
    \right]- E_0,
    \label{Eq:Energy}
\end{equation}
with $j$ the vorticity index. $E_0$ is the energy of the homogeneous condensate without vortices.
Finally, the interaction energy of two vortices 
is $E_{int}  = E_2  - 2E_1 $, where $E_2$ is the energy for two vortices separated by $r_{vv}= r_1-r_2$ (with $\Psi_2(\mathbf{\mathbf{r}})$ given by Eq.\ \eqref{Eq:2Vwavefunction}), and $E_1$ is the energy with just one vortex (with $\Psi_1(\mathbf{r})$ given by Eq.\ \eqref{Eq:1Vwavefunction}).

Figure \ref{fig:energy} shows the calculated vortex-vortex interaction energy $E_{int}$ for two vortices at separation $r_{vv}= r_1-r_2$, at fixed density and for different interlayer distances $d$. 
The vortex-vortex interaction is everywhere repulsive, i.e. $E_{int}>0$ \cite{Mulkerin2013}. 

For large inter-vortex distance $r_{vv}$, the $E_{int}$ all follow the general behavior for vortex-vortex interactions in superfluids  \cite{Pethick2008, Martin2017}.
However, at short range ($r_{vv}<25$) there are significant deviations from this behavior that become more and more striking with increasing interlayer spacing $d$ (i.e. with increasing exciton dipole moments). These arise because the $\phi_{\text{XX}}$ contribution in Eq.\ \eqref{Eq:Energy} starts to dominate. 
This narrows the vortex core and reduces the core overlap. 
For small $r_{vv}<2R_c$, the cores overlap and the energy cost to create two single vortices equals the energy cost to create a ``giant'' vortex with vorticity $\ell=2$. This causes the saturation of $E_{int}$ for small $r_{vv}$.

\begin{figure*}[t]
\includegraphics[width=\textwidth]{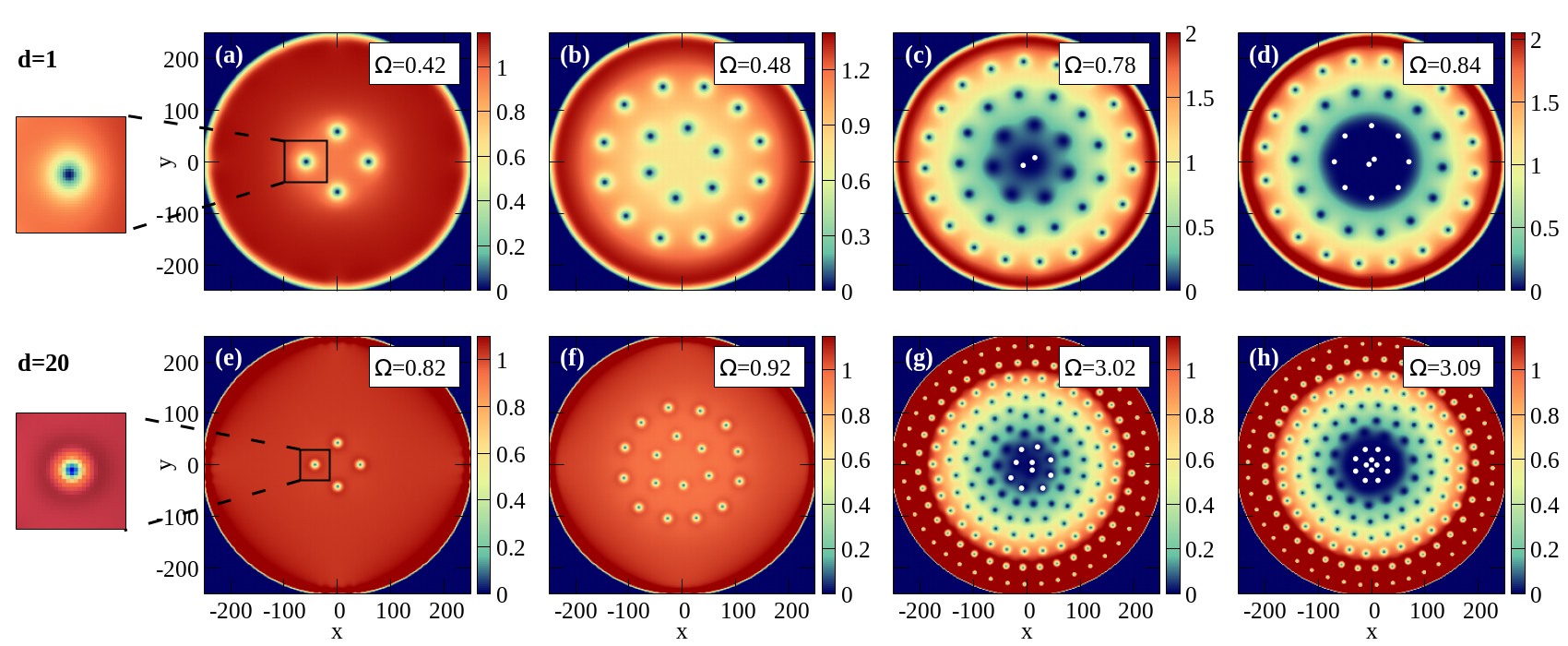}
\caption{Contour-plots of the exciton superfluid density distribution $|\Psi(\mathbf{r})|^2$ for $r_0=30$, normalized to the homogeneous density. 
(a-d) for $d=1$, and (e-h) for $d=20$, with different rotation frequencies $\Omega$ in mRy$^*\!\!/\hbar$ as indicated in each panel.
A magnified view of a vortex from panels (a) and (e) is shown on extreme left.
The white dots in selected panels mark the centers of the strongly overlapping vortex cores, as determined by the singularities in the phase of the order parameter (see Fig.\ \ref{fig:phase}).
}
\label{fig:VL}
\end{figure*}

\begin{figure}[t]
\includegraphics[width=\columnwidth]{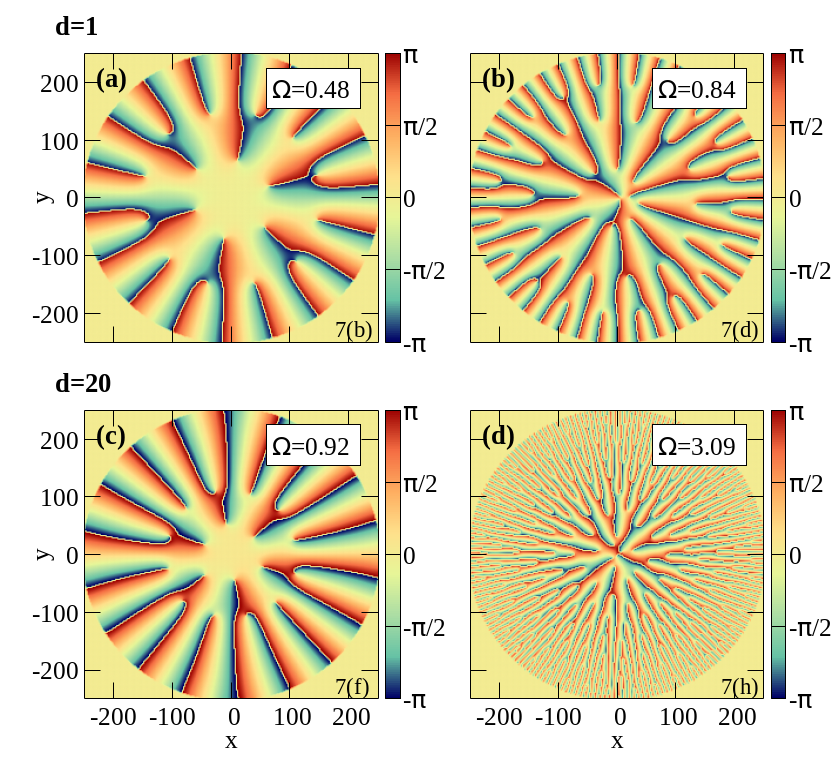}
\caption{
Contour-plots of the phase of $\Psi(\mathbf{r})$ for $r_0=30$, with $d$ and $\Omega$ as indicated.  
Panels (a)-(d) correspond to selected states from Fig.\ \ref{fig:VL} as labeled in the bottom right corners.}
\label{fig:phase}
\end{figure}
\begin{figure}[t]
\includegraphics[width=\columnwidth]{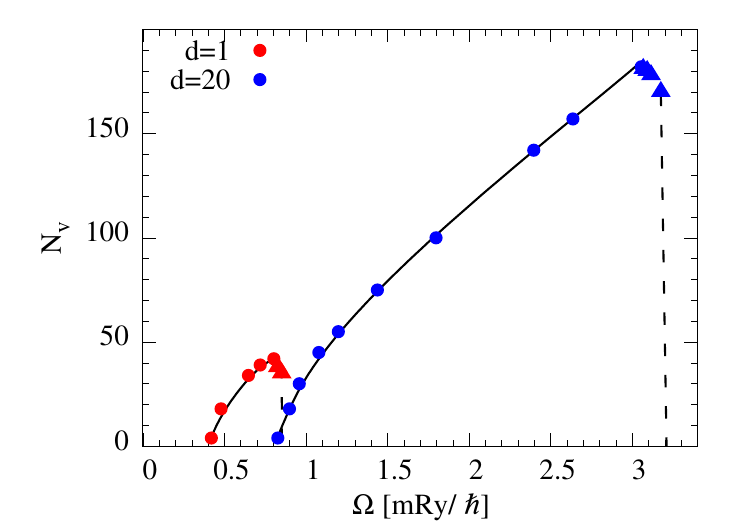}
\caption{Number of distinguishable vortices $N_v$ in the vortex lattice as a function of the rotation frequency $\Omega$ for two interlayer separations $d$. Dots mark the states where all $\ell=1$ vortices are clearly observable in the density. Triangles mark the states where clusters of strongly overlapping vortex cores form in the center of the condensate, and only vortices outside the cluster are distinguishable in the density profile.  Solid lines are guides for the eye.}
\label{fig:NVL}
\end{figure}

\begin{figure}[t]
\includegraphics[width=\columnwidth]{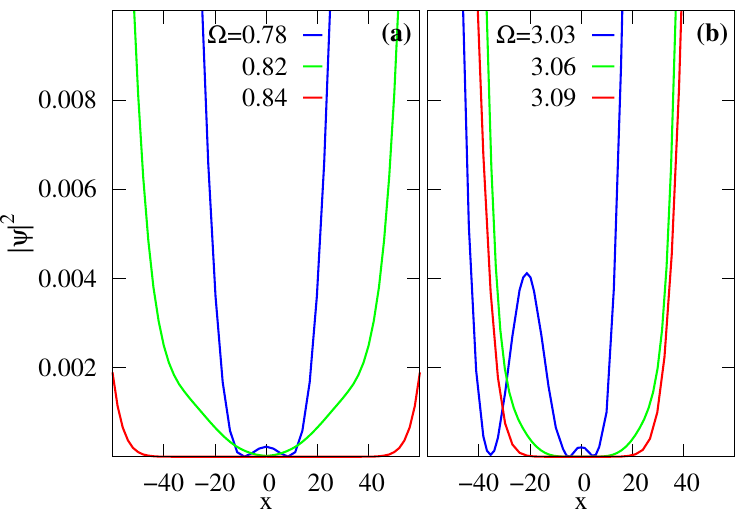}
\caption{Line profile along the $x$-axis of the exciton superfluid density distribution $|\Psi(\mathbf{r})|^2$ for $r_0=30$, normalized to the homogeneous density, for (a) $d=1$ and (b) $d=20$.}
\label{fig:profile}
\end{figure}

\subsection{Vortex lattice}
\label{Sec:Lattice}

In this section, we generalize Secs.\ \ref{Sec:SingleVortex}-\ref{Sec:VortexInt} for a realistic experimental configuration in which the vortices are generated by imprinting a rotational momentum on the system.
The Hamiltonian in a uniformly rotating frame with angular velocity $\mathbf{\Omega}$, leads to the Gross-Pitaevskii equation in the form:
\begin{equation}
-\frac{\hbar^2}{2M} \nabla^2 \; \Psi(\mathbf{r}) +  \phi_{\text{XX}}(\mathbf{r})\Psi(\mathbf{r}) -\mathbf{\Omega} \cdot \mathbf{L} \, \Psi(\mathbf{r}) =\mu \Psi(\mathbf{r}),
\label{Eq:DGPEquationROT}
\end{equation}
where  $\mathbf{L}=-i\hbar \,\mathbf{r} \times \nabla$ is the angular momentum. 
In the simulations the sample has radius $250$.
We start from a randomized order parameter and iterate the solution up to a convergence criterion as described in Section \ref{Sec:Theoretical}. 
After finding a stable solution, we progressively increase angular momentum and repeat the procedure, without keeping history of the previously found state.

Above a critical angular velocity $\Omega_c$, vortices enter the system, forming a stable vortex lattice. 
Figure \ref{fig:VL} shows the superfluid density profile $|\Psi(\mathbf{r})|^2$  of representative vortex lattices for $d=1$ and $20$. 
Fig.\ \ref{fig:phase} showing the corresponding phase of $\Psi(\mathbf{r})$ for selected panels.  
The circulation in Fig.\ \ref{fig:phase} shows that the vortices enter from the edges as single vortices, all with the same circulation from $-\pi$ to $\pi$.  This is in contrast with other systems like superconductors, where the vortices enter from the center as vortex-antivortex pairs.

Similarly to the single-vortex discussion in Sec.~\ref{Sec:SingleVortex},
Fig.\ \ref{fig:VL} shows the shrinking of the radii of the vortex cores when $d$ is increased and for $d=20$, the density pileup peak is clearly visible at the edge of the vortex cores (see magnified views to the left of Figs.\ \ref{fig:VL}(a) and (e)).  

The critical $\Omega_c$ and the $\Omega$ needed to achieve the same vorticity both increase with interlayer separation $d$ as is seen in  
Fig.\ \ref{fig:NVL}, which shows the number of distinguishable vortices $N_v$ in the lattice as a function of $\Omega$, determined as distinguishable zeros in the order parameter.
This is due to the increased strength of the exciton interaction energy \cite{Correggi2012,Martin2017}.
Figures \ref{fig:VL}(a),(b) and (e),(f) show systems with relatively few vortices in stable lattices with the same vorticity.
The vortex separations are large, $r_{vv}>30$, and so the vortex-vortex repulsion is the same for $d=1$ and $20$ (see Fig.\ \ref{fig:energy}).
The larger $\Omega$ for $d=20$ means that its rotation energy, $\frac{1}{2}\Omega^2 r^2$~\cite{Correggi2013}, is larger than for the system with $d=1$. 
Thus to minimize the energy, the vortices are pushed closer together for $d=20$ than for $d=1$. 

As $\Omega$ increases, the number of vortices increases and the vortex separation decreases. 
Figures \ref{fig:VL}(c) and (g) show the point at which the vortex cores at the center of the condensate start to overlap. 
This is clearly visible from the blue curves in Fig.\ \ref{fig:profile} showing a cut along the $x$-axis of the $|\Psi(\mathbf{r})|^2$ at the center of the sample. 
Further increasing $\Omega$ leads to a merger of vortex cores (green lines in Fig.\ \ref{fig:profile}). 
Figures \ref{fig:VL}(d) and (h) show that at still larger $\Omega$, an increasing number of vortices accumulate in a cluster of overlapping vortices at the center of the condensate.
This occurs in a region where the order parameter is strongly suppressed, extending across the entire cluster (red lines in Fig.\ \ref{fig:profile}). 
Figure \ref{fig:profile} shows that the width of the cluster decreases with increasing $d$.
The formation of the cluster is intimately connected to the vortex separations $r_{vv}$ falling below the value at which the vortex-vortex interaction energy first saturates (see Fig.\ \ref{fig:energy}).

In Fig.\ \ref{fig:NVL}, the triangles mark the states where these clusters of strongly-overlapping vortex cores form. 
As a consequence, the total number of distinguishable vortices decreases, since the only vortices outside the cluster are distinguishable in the density profile.
$|\Psi(\mathbf{r})|$ is indistinguishably suppressed within the cluster, but nevertheless the topological singularities of each vortex are still identifiable from the phase of $\Psi(\mathbf{r})$. 
Figures \ref{fig:phase}(b) and (d) show that the phase singularities in the center of the sample in fact do not merge.
This indicates that the single vortices retain their identity in the cluster.

Finally, for still larger $\Omega$, Fig.\ \ref{fig:NVL} shows that the system no longer supports formation of a vortex lattice.  We find that the condensate collapses from inside out~\cite{Zwierlein2005}.

\section{Conclusions}
\label{Sec:Conclusions}
We have characterized vortices in the superfluid state of an exciton bilayer system using the Gross-Pitaevskii formalism for non-local interactions.
The exciton dipolar-like interactions are purely repulsive, leading to significantly different properties as compared to cold atom systems.   

We found that vortex properties can be tuned by modifying the strength of the exciton-exciton repulsion through variation of the interlayer distance (exciton dipole moment) and exciton density. 
The core radius of the vortices shrinks when $d$ is increased, and for sufficiently large $d$, the radius saturates. Simultaneously, a pileup of the density appears at the edge of the vortex. 
Both these effects are due to strong exciton-exciton repulsion which at large $d$ eventually dominates over the outward-directed centrifugal force, thus preventing the vortex from expanding. 
A pileup peak at the edge of the vortices has also been reported in dipolar cold atoms, but unlike with cold atoms, here there are no subsequent density oscillations accompanying the peak.  
The absence of oscillations in our system suggests there is no low-lying roton excitation, which is likely due to the purely repulsive nature of the aligned-exciton interactions.

We also show that the vortex-vortex interactions are tuned by the strength of the exciton-exciton interactions via the interlayer spacing $d$. 
The vortex-vortex interactions increase as the vortices get closer and this increase gets larger as $d$ grows. 
We find that the cores overlap at very small inter-vortex distances less than the vortex diameter.  
Then two single vortices can cost more energy to create than one ``giant'' vortex with double the vorticity.  
This leads to a saturation of the interaction energy.

Finally, for the time-independent Gross-Pitaevskii equation in a rotating frame, we have followed the formation of a stable vortex lattice. 
We confirm these stable solutions show the same dependence of the vortex core size on the strength of the exciton-exciton interaction and, for strong interactions, we confirm a similar appearance of a density pileup at the edge of the vortices -- as we saw in the single vortex case. 
We characterize the vortex lattices for varying rotational frequency $\Omega$. 
We show that the critical $\Omega_c$ for the initial appearance of vortices depends on the exciton interaction strength -- through the interlayer separation $d$. 
With increasing $\Omega$, the system traverses a number of different regimes \cite{Correggi2013}: (i) a lattice of well separated vortices; (ii) a lattice with a steadily increasing vortex overlap at its center; (iii) a lattice with a cluster of strongly overlapping vortices in its interior; and (iv) the condensate collapse from inside out.

An interesting apparent link emerges from the results, since the appearance of the density pileup peak, accompanied by a saturation of the vortex core radius, first occurs at very similar values of $d$ as those predicted for the superfluid to incompressible supersolid transition \citep{Conti2023}. 
At the transition, the vortex would be sufficiently compact to fully fit within a unit cell of the incompressible supersolid.
Besides the appearance of vortices serving as conclusive 
evidence for the existence of exciton superfluidity, the possibility of an intriguing link of the observable vortex properties to the incompressible supersolid transition clearly opens exciting experimental opportunities.

\section*{Acknowledgements}
We thank Jacques Tempere for useful discussions.
This work was supported by Research Foundation -- Flanders (FWO-Vl) and by the Brazilian Council for Research (CNPq), through the PRONEX/FUNCAP, Universal, and PQ programs. S.C. is a senior postdoctoral fellow of FWO-Vl, under contract No. 12A3T24N.


%

\end{document}